\documentclass
[a4paper,10pt,reprint,longbibliography,aps,superscriptaddress,pra,
amsmath, amssymb, amsfonts]
{revtex4-1}
\usepackage{comment} % for longer comments
\usepackage{siunitx} % standard-conform number-unit distance and typeset
\DeclareSIUnit \Uc{\mathit{U}_\mathrm{cl}}
\DeclareSIUnit \Lc{\mathit{R}}
\usepackage{graphicx,epstopdf} % to include figures
\usepackage[svgnames]{xcolor} % way more color options with svgnames 
\usepackage{tabularx,booktabs,dcolumn,multirow}%,multicol} % better tables
\usepackage{bm} % bm is a better alternative to \boldsymbol or \bf for
\usepackage{empheq} % grouping eqns.
\usepackage[abs]{overpic} % lifesaver!

\newcommand{\clr}[2]{\textcolor{#1}{#2}} % for comments; #1: color, #2:text
\newcommand{\etal}{\mbox{\emph{et al.}}\ }

\newcommand{\ie}{\mbox{\emph{i.e.}\ }}
\newcommand{\D}{\mathrm{D}} % capital differential
\newcommand{\pp}[2]{\frac{\partial #1}{\partial #2}} % partial derivative
\newcommand{\DD}[2]{\frac{\D #1}{\D #2}} % material derivative
\newcommand{\vc}[1]{\bm{#1}} % Uncomment for BOLD vectors.
\newcommand{\R}{\textrm{Re}} % Reynolds number shortcut #1
\newcommand{\RE}[1]{\textrm{Re}=#1} % Reynolds number shortcut #2

\graphicspath{{./fig/}}

\begin{document}

   \title{\clr{DodgerBlue}
         {Analysis and modeling of localized invariant solutions in  
          pipe flow}
   }

   \author{Paul Ritter}
   \email[Corresponding author: ]{paul.ritter@zarm.uni-bremen.de}
   \affiliation{
   Center of Applied Space Technology and Microgravity (ZARM), 
   University of Bremen, 28359 Bremen, Germany}
   \affiliation{
   Institute of Fluid Mechanics, Friedrich-Alexander-Universit\"at
   Erlangen-N\"urnberg, 91058 Erlangen, Germany}

   \author{Stefan Zammert}
   \affiliation{
   Laboratory for Aero- and Hydrodynamics, TU Delft, 2682 Delft, The
   Netherlands}

   \author{Bruno Eckhardt}
   \affiliation{
   Fachbereich Physik, Philipps-Universit\"at Marburg, 35032 Marburg,
   Germany}
   \affiliation{
   JM Burgerscentrum, TU Delft, 2682 Delft, The Netherlands}

   \author{Marc Avila}
   \affiliation{
   Center of Applied Space Technology and Microgravity (ZARM), 
   University of Bremen, 28359 Bremen, Germany}
   \affiliation{
   Institute of Fluid Mechanics, Friedrich-Alexander-Universit\"at
   Erlangen-N\"urnberg, 91058 Erlangen, Germany}
   
   \date{\today}

   \begin{abstract}
     %!TEX root=../main.tex
Turbulent spots surrounded by laminar flow are a landmark of
transitional shear flows, but the dependence of their kinematic
properties on spatial structure is poorly understood.  We here investigate 
this dependence in pipe flow for Reynolds numbers between 1500 and 5000. We 
compute spatially localized relative periodic orbits in long pipes and show 
that their upstream and downstream fronts decay exponentially towards the 
laminar profile.  This allows to model the fronts by employing the 
linearized Navier--Stokes equations, and the resulting model yields
the spatial decay rate and the front velocity profiles of the 
periodic orbits as a function of Reynolds number, azimuthal wave number and 
propagation speed. In addition, when applied to  a localized turbulent 
puff, the model is  shown to accurately approximate the spatial decay rate of 
its upstream and downstream tails. Our study provides insight into the 
relationship between the kinematics and spatial structure of localized 
turbulence and more generally into the physics of localization.
   \end{abstract}

   \maketitle 

   %!TEX root=../main.tex 
\section{Introduction} 

Due to its stochastic and fluctuating nature, the classical approach 
towards understanding turbulent fluids has been a statistical one, which 
dates back to Osbourne Reynolds \cite{rans}. In recent years an alternative 
approach has emerged, in which the (discretised) Navier-Stokes equations 
are viewed as a high-dimensional dynamical system and the tools of 
bifurcation and chaos theory are applied to describe turbulent motions 
\cite{kerswell2005,eckhardt2007,eckhardt2008}. The key idea of this approach
is that the turbulent dynamics is shaped by simple exact invariant solutions
to the governing equations such as traveling waves \cite{faisst2003,
wedin2004} and relative periodic orbits \cite{duguet2008, willis2013} in 
pipe flow. The dynamically most relevant solutions have a relatively small
number of unstable directions so that a generic turbulent trajectory spends
a significant amount of time in their vicinity \cite{gibson2008, 
kawahara2012, schneider2007b}. Turbulent trajectories depart
from the vicinity of the solutions along their unstable manifolds, which can subsequently govern the flow
evolution for a considerable period of time \cite{suri2017}. 
In principle, all properties of the turbulent flow can be derived by an appropriate weighted average over the 
fundamental solutions \cite{predrag1988, predrag1989, predrag1991, 
predrag2013, kreilos2012}, but deploying this approach is extremely
challenging even at low (transitional) Reynolds numbers \cite
{schneider2007b, willis2016, kreilos2012}.

Transitional shear flows are characterized by localized chaotic spots 
surrounded by laminar flow \cite{reynolds1883, emmons1951, tillmark1992, 
lemoult2013}. These spots already contain all the salient features of 
fully turbulent flow \cite{wygnanski1973,barkley2015,cerbus2017} and hence pose an 
ideal prototype for a bottom-up study of turbulent dynamics. Because of 
prevalence of intermittency at the onset of turbulent shear flow 
\cite{rotta1956, moxey2010, avila2011}, spatially localized invariant 
solutions are indispensable for its successful description as a dynamical 
system. The first such solutions were discovered in plane Couette flow by 
Schneider \etal\cite{schneider2010}, who computed spanwise-localized 
equilibria and traveling waves in wide but streamwise short domains. 

\begin{figure} 
   \begin{overpic}[width=\linewidth,keepaspectratio] 
      {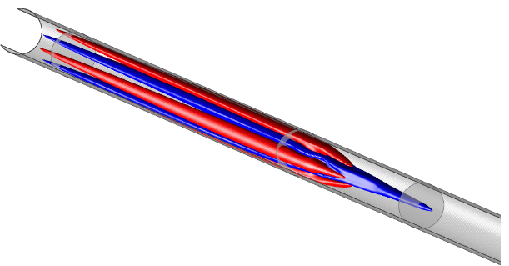} 
      \put(0,135){(a)} 
      \put(96,90)
      {\includegraphics[width=0.6\linewidth]{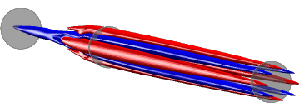}} 
      \put(170,127){back view} 
      \put(20,28)
      {\includegraphics[width=0.2\linewidth]{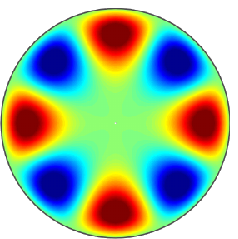}} 
      \put(27,18){upstream}
      \put(100,-7)
      {\includegraphics[width=0.2\linewidth]{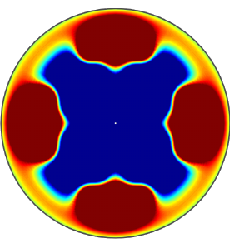}} 
      \put(117,-17){core}
      \put(180,-42)
      {\includegraphics[width=0.2\linewidth]{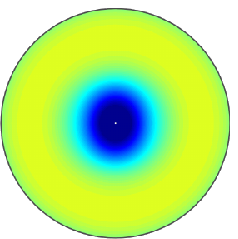}} 
      \put(182,-52){downstream}
   \end{overpic}\vspace{30mm} 
   \begin{overpic}[width=\linewidth,keepaspectratio] 
      {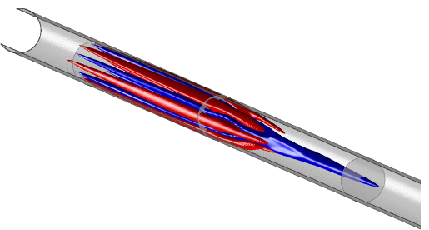} 
      \put(0,135){(b)} 
      \put(100,92)
      {\includegraphics[width=0.55\linewidth]{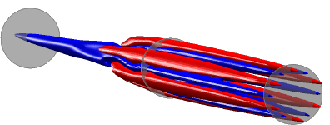}} 
      % \put(150,80){back view} 
      \put(20,28)
      {\includegraphics[width=0.2\linewidth]{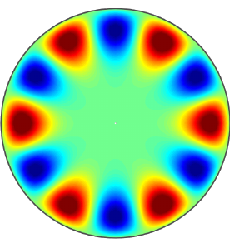}} 
      \put(100,-7)
      {\includegraphics[width=0.2\linewidth]{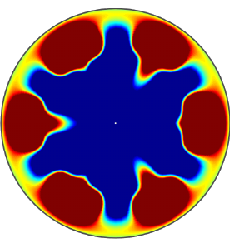}} 
      \put(180,-42)
      {\includegraphics[width=0.2\linewidth]{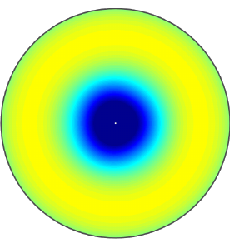}} 
   \end{overpic}\vspace{14mm} 
\caption{
   Structure of the investigated localized reflection-symmetric relative
   periodic solutions obtained by directly solving the Navier--Stokes
   equation with a Newton--Krylov method at $\RE{3000}$, (a) twofold
   (LB$_2$), (b) threefold (LB$_3$) rotational symmetry. The central image
   shows isosurfaces and cross-sections of the streamwise velocity
   disturbance (\ie with the laminar flow subtracted). Red (blue) streaks are
   faster (slower) than base flow. In order to highlight the tails of the
   solutions, the isovalues have been chosen small: $\pm$\,\SI{0.025}{\Uc}.
   The axial extent of the shown isosurfaces is $\approx$ \SI{60}{\Lc}. The
   upper right panel shows the upstream front of the same state, which
   appears shorter due to a different perspective.
} 
\label{fig:iso_sec} 
\end{figure}

The first streamwise-localized simple invariant solution was found by Avila
\etal in pipe flow \cite{avila2013}. It is a relative periodic orbit with 
reflectional and two-fold rotational symmetry appearing at a saddle-node
bifurcation. In this symmetry subspace the lower branch solution (shown in
fig.~\ref{fig:iso_sec}a and referred to as LB$_2$ in the following) has a
single unstable direction, whereas the upper branch solution is stable close
to the saddle-node bifurcation. As the Reynolds number increases, a
bifurcation cascade culminating at a boundary crisis gives rise to transient
chaotic dynamics \cite{avila2013,ritter2016}, and subsequent changes in the
phase-space progressively enhance the spatio-temporal complexity of the flow
\cite{fer2009, ritter2016}. The bifurcations of the coherent structures in
pipe flow follow the same pattern as observed in small computational cells in plane Couette  \cite
{kreilos2012, kreilos2014} and plane Poiseuille flows \cite{zammert2015}. The
finding of spanwise- and streamwise-localized solutions in both flows
\cite{zammert2014, gibson2014} suggests that similar scenarios may occur
in spatially extended domains.

Gibson and Brand \cite{gibson2014} observed that the amplitude of spanwise 
localized equilibria in Couette flow decays exponentially far enough from 
their energetic core. Hence they proposed to model their spatial decay by 
using the linearized Navier--Stokes equations and solving the arising 
eigenvalue problem. Interestingly, their model gave with high accuracy the observed 
spatial decay rates and their dependence on streamwise wave number and the
Reynolds number. A simplified version of their model was shown by the same 
authors to accurately reproduce the spatial decay of their doubly-localized solutions in 
the streamwise direction \cite{brand2014}. 

Recently, Zammert and Eckhardt \cite{zammert2016} and Barnett \etal \cite
{grigoriev2016} extended this approach to streamwise relative periodic orbits 
in channel flow, where the decay rates also depend on their group velocity,
in addition to the spanwise wave number and the Reynolds number. In this
paper, we use a similar approach to pipe flow. For this purpose, we compute
LB$_2$ and its three-fold cousin \cite{chantry2014} (LB$_3$, see fig.~\ref
{fig:iso_sec}b) for a wide range of Reynolds numbers. As in channel flows, we
find that the tails of the states decay exponentially in the streamwise
direction and that the decay rates can be deduced from the linearized 
Navier--Stokes equations. However, pipe flow does not permit the simplified
modeling approaches used by Brand and Gibson \cite{brand2014}, and Zammert
and Eckhardt \cite{zammert2016}. Furthermore, we extend the analysis to the
upstream and downstream tails of localized turbulent puffs and verify that
the correct decay rates are obtained for such chaotically evolving states as
well.
   %!TEX root=../main.tex
\section{Numerical method}\label{sec:methods}

We consider the incompressible, isothermal flow of a fluid with
constant density $\rho$ and kinematic viscosity $\nu$ in a cylindrical
pipe of radius $R$ driven at a constant average speed
$\overline{U}$. This flow is governed by the Navier-Stokes equations
(NSE):
\begin{subequations}\label{eq:ns}
   \begin{empheq}{align}
      &\DD{\vc{U}}{t} = \pp{\vc{U}}{t} + \vc{U} \cdot
     \vc{\nabla} \vc{U} = -\dfrac{1}{\rho}\vc{\nabla} P +
     \nu\nabla^2\vc{U}\\ & \vc{\nabla} \cdot \vc{U}
     =\left(\frac1{r}+\pp{}{r}\right)U_r +
     \frac1{r}\pp{}{\theta}U_{\theta} + \pp{}{z}U_z = 0,
   \end{empheq}
\end{subequations}
where $P$ is the pressure and $\vc{U} = [U_r,U_\theta,U_z](r,\theta,z,t)$ is
the fluid velocity field in cylindrical coordinates. It satisfies the no-slip 
boundary condition at the pipe wall and periodic
boundary conditions in the azimuthal and axial directions. The length of the
computational domain was chosen sufficiently large in order to avoid
interaction of the two fronts via the axial periodicity. All results shown
in this paper were obtained in pipes of \SI{200}{\Lc} (LB$_2$) and \SI{160}
{\Lc} (LB$_3$) in length.

The \emph{Hagen-Poiseuille} profile is the steady, parabolic laminar
solution and reads (subscript ``b'' is for \emph{base flow}):
\begin{align} \vc{U}_\mathrm{b} \quad &= \quad
    U_\mathrm{cl}\left[1-\left(\frac{r} {R}\right)^2\right]\vc{\hat{z}},
    \quad
    U_\mathrm{cl}=2\,\overline{U},\\ -\vc{\nabla} P_\mathrm{b} \quad &= \quad
    \Pi_\mathrm{b}\vc{\hat{z}}=
   \frac{4\mu U_\mathrm{cl}}{R^2}\,\vc{\hat{z}},
\end{align}
where $U_\mathrm{cl}$ is the maximum velocity at the centerline, and
$\vc{\hat{z}}$ denotes the axial unit vector. To facilitate both numerical
and theoretical treatment, the NSE are rendered dimensionless by using
$U_\mathrm{cl}$, $\rho U_\mathrm{cl}^2$ and $R$ as reference scales for the
velocity, pressure and length, respectively. As a consequence, the
dimensionless NSE are identical to eq.~\eqref{eq:ns} but setting $\rho=1$
and replacing the viscosity with the inverse of the Reynolds number
$1/\R=\nu/(U_\mathrm{cl}R)$, which is the sole control parameter of the
problem. The velocity and pressure gradient of the dimensionless laminar flow
then take the form $(1-r^2)\vc{\hat{z}}$ and $4/\R$, respectively. Throughout
the paper the velocity disturbance $\vc{u}=\vc{U}-\vc{U}_\mathrm{b}$ is used
to visualize the structures.

The direct numerical simulations of the Navier--Stokes equations
\eqref{eq:ns} have been carried out using
\href{http://openpipeflow.org}{openpipeflow.org} \cite{openpipeflow},
a hybrid spectral finite-difference Navier-Stokes solver, which uses
primitive variables and a PPE-formulation with correct pressure boundary
conditions via the influence-matrix method \cite{rempfer2006, guseva2016}.
In order to compute the localized structures, a two-step approach was
employed. First, the edge-tracking technique \cite{itano2001, skufca2006}
was used to bracket the relative periodic orbits to a sufficient degree so
that it could be converged in a second step with a Newton--Krylov-hookstep
algorithm \cite{viswanath2007} to relative error $10^{-6}$. The necessary
spatial resolution of the periodic directions depends on the enforced
rotational symmetry. We used an axial resolution $-K\ldots K$ of $\pm\,768$
Fourier modes for a pipe of length \SI{200}{\Lc} in case of two-fold
symmetry and the same amount of modes for a \SI{160}{\Lc} pipe in the
three-fold case. The spanwise resolution was $\pm\,12$
($\pm\,16$) Fourier modes for LB$_3$ (LB$_2$) capturing up to 36th (32nd)
wave number (of which a third/half has the same amplitude due to symmetry).
The code uses the $3/2$-rule for dealiasing (\ie padding) resulting in a
physical grid which has three times as many points as there are wave numbers
$K$ ($3/2*2K$). The radial direction has been discretized with a minimum of $48$ and maximum of $72$
finite-difference nodes depending on $\R$ and pipe length. The time step was
fixed at a value of \SI {0.01}{\Lc/\Uc}.
   %!TEX root=../main.tex
\section{Exponential localisation of solutions}\label{sec:results}

\begin{figure}\vspace{-1mm} % raise a bit bec. ax. label/(a) sticks out
   \begin{tabularx}{\linewidth}{c}
      \begin{overpic}[width=\linewidth]{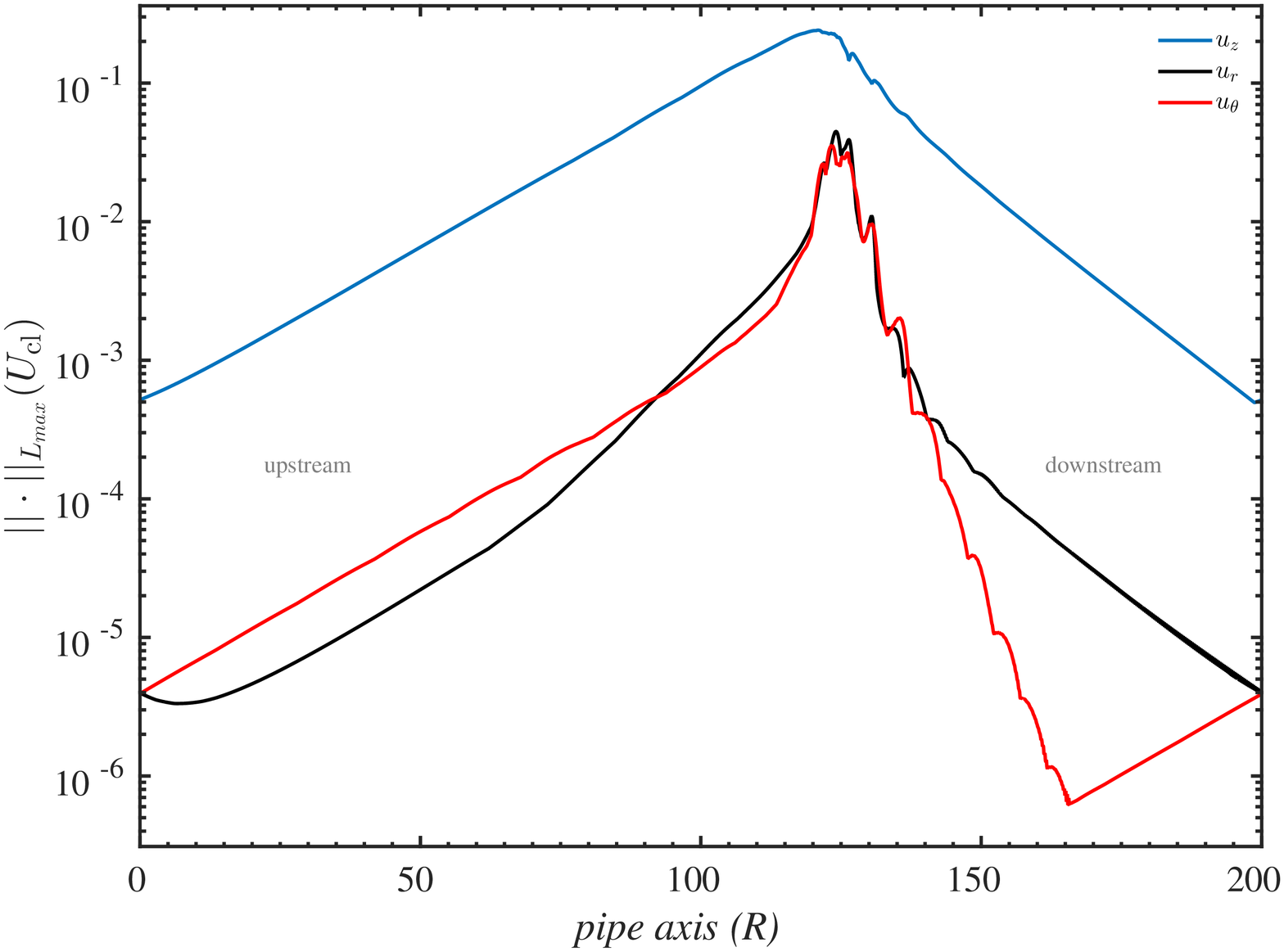}
         \put(-5,178){(a)}
      \end{overpic} \\
      \begin{overpic}[width=\linewidth]{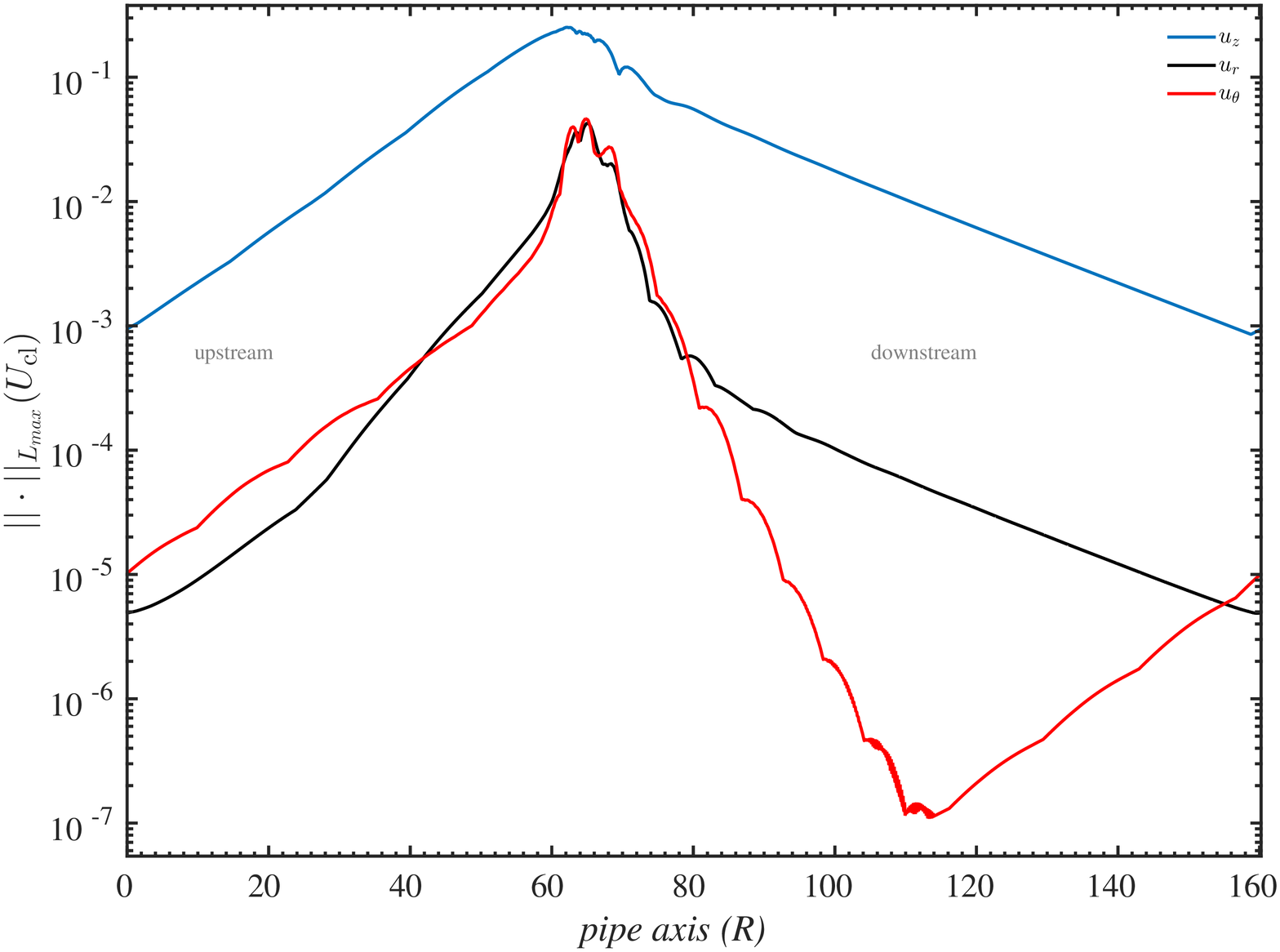}
         \put(-5,178){(b)}
      \end{overpic}
   \end{tabularx}
\caption{Axial profiles of the infinity norm of the disturbance velocity
components at $\RE{3000}$,  (a) LB$_2$. (b) LB$_3$.}
\label{fig:norm_lengths}
\end{figure}

\begin{figure}%\vspace{1mm}
   \begin{overpic}[width=\linewidth,keepaspectratio]
      {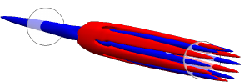}
      \put(0,75){(a)}
      \put(23,28){downstream}
      \put(20,-28){\includegraphics[width=0.2\linewidth]{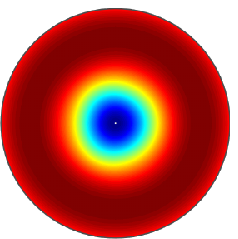}}
      \put(187.2,-9){upstream}
      \put(180,-65){\includegraphics[width=0.2\linewidth]{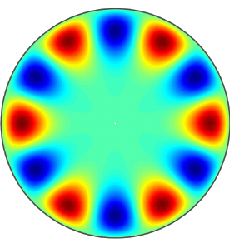}}
   \end{overpic}\vspace{23mm}
   \begin{overpic}[width=\linewidth,keepaspectratio]
      {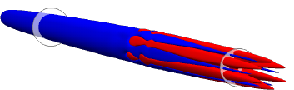}
      \put(0,75){(b)}
      \put(20,-23){\includegraphics[width=0.2\linewidth]{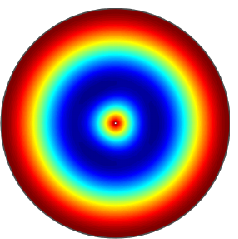}}
      \put(180,-55){\includegraphics[width=0.2\linewidth]{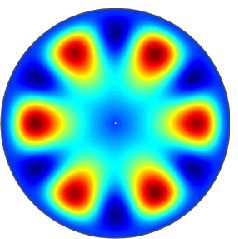}}
   \end{overpic}\vspace{23mm}
   \begin{overpic}[width=\linewidth,keepaspectratio]
   {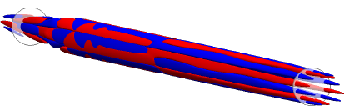}
      \put(0,75){(c)}
      \put(20,-23){\includegraphics[width=0.2\linewidth]{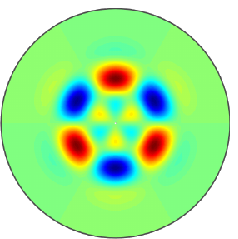}}
      \put(180,-55){\includegraphics[width=0.2\linewidth]{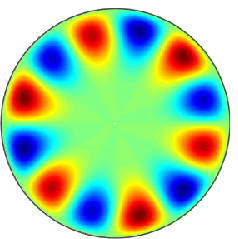}}
   \end{overpic}\vspace{18mm}
\caption{
Comparison of the spatial structure of the three velocity
components for LB$_3$ at $\RE{3000}$. The color coding is analogous to
fig.~\ref{fig:iso_sec} and the isovalues are given in parantheses.
(a) $u_z$ ($\pm$\,\SI{0.015}{\Uc}), (b) $u_r$ ($\pm$\,\SI{0.5e-4}{\Uc}), (c)
$u_{\theta}$ ($\pm$\,\SI{0.5e-5}{\Uc}).
}
\label{fig:structure_fronts}
\end{figure}

The approach of the fields to the asymptotic parabolic flow is best
visualized by the deviation $\vc{u}$ from the Hagen-Poiseuille profile,
since they have to decay to zero. The isosurfaces of streamwise velocity
deviation from laminar flow $u_z$ shown in fig.~\ref{fig:iso_sec} illustrate
the spatial arrangement of streaks of the spatially localized relative
periodic orbits LB$_2$ and LB$_3$ at $\R =3000$. Far from the active core,
all three velocity components decay quickly with respect to the streamwise
direction $z$. The semilogarithmic representation in fig.~\ref
{fig:norm_lengths} shows that the decay is predominantly exponential, with
$u_z$ approximately two orders of magnitude larger than the cross-stream
velocities $u_r$ and $u_\theta$, thus dominating the decay toward
laminar flow. Interestingly, the decay rate of the azimuthal velocity at the
downstream tail differs from that of the other two components. To shed light
on the origin of this difference, isosurfaces of all three velocity
components are shown in fig.~\ref{fig:structure_fronts} for LB$_3$. In the
upstream tail all three velocity components feature a predominant sixfold
rotational symmetry, whereas in the downstream tail $u_r$ and $u_z$ are
predominantly axisymmetric and $u_\theta$ features a threefold symmetric
structure. LB$_2$ exhibits the same features but with fourfold and twofold
symmetry, instead of sixfold and threefold, respectively, and hence it is
not shown here.

\begin{figure}
   \begin{tabularx}{\linewidth}{c}
      \begin{overpic}[width=\linewidth]{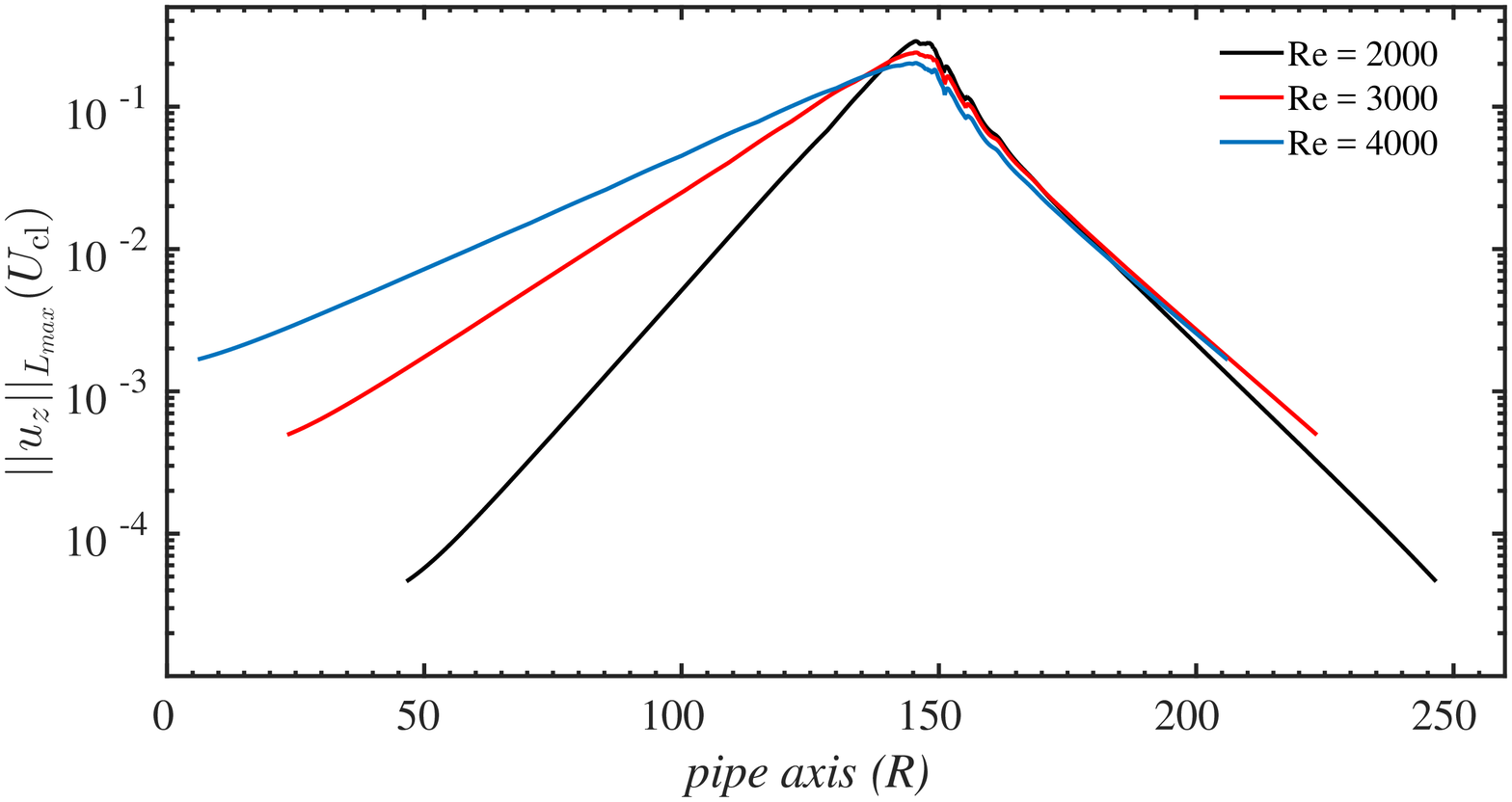}
         \put(-5,126){(a)}
      \end{overpic} \\
      \begin{overpic}[width=\linewidth]{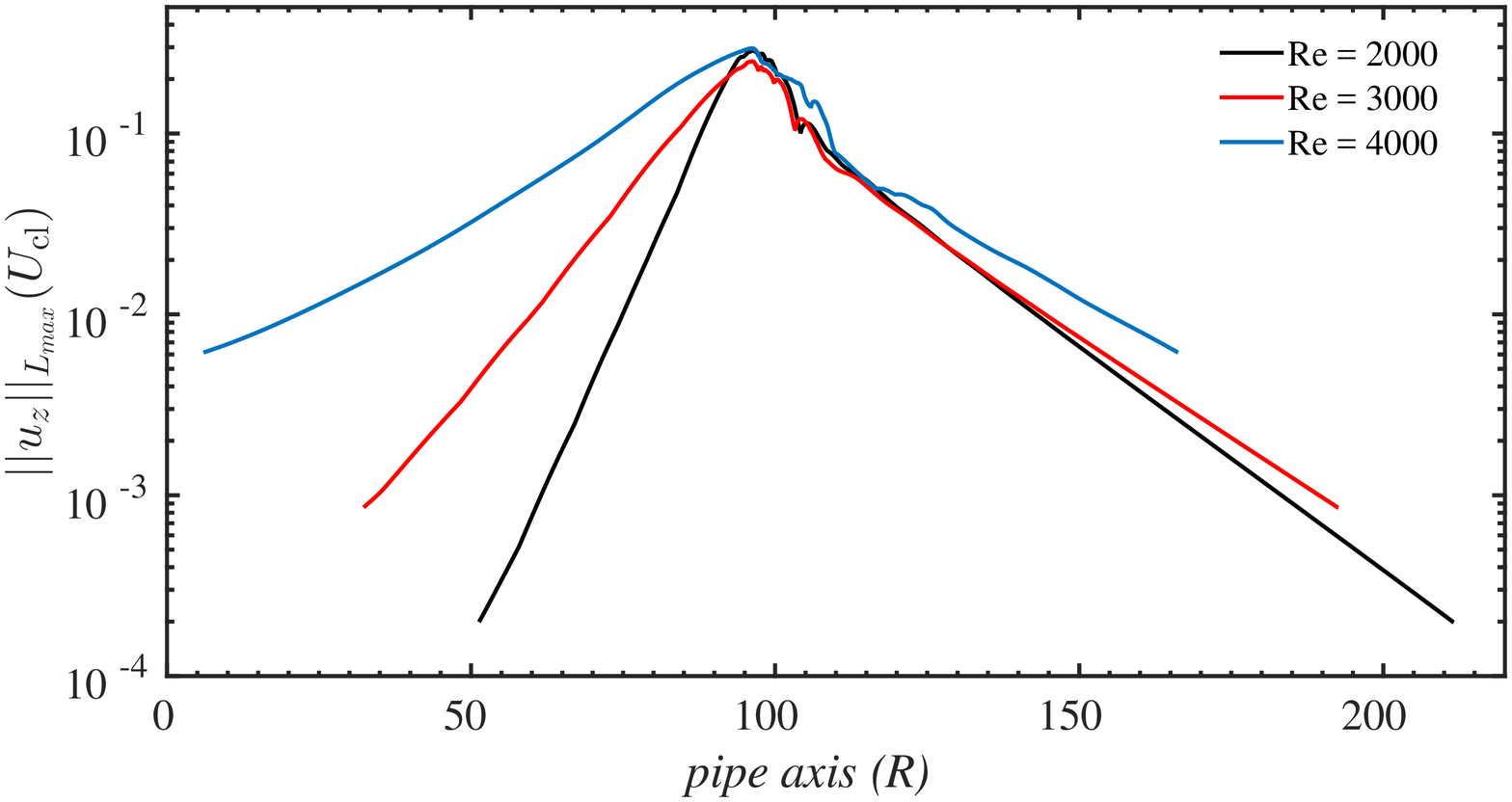}
         \put(-5,126){(b)}
      \end{overpic}
   \end{tabularx}
\caption{Axial profiles of the infinity norm of the axial velocity
disturbance for different Reynolds numbers. (a) LB$_2$. (b) LB$_3$.}
\label{fig:norm_Res}
\end{figure}

The length of the core of LB$_2$ and LB$_3$ remains nearly constant, whereas
its amplitude decreases as \R\ increases (see fig.~\ref{fig:norm_Res}). 
This is not surprising because LB$_2$ and LB$_3$ are edge states and can 
thus be seen as minimal seeds to trigger turbulence \cite{pringle2012}. The decay rate of both 
their upstream and downstream tails decreases with \R, \ie the axial velocity 
profile gradually ``opens up''. There is a marked asymmetry, however. While
the decay rates of the downstream tails change little with $\R$, the decay
rates of the upstream tails decrease rapidly with increasing $\R$. This 
behaviour is similar to that for the relative periodic orbits in plane Poiseuille
flow \cite{zammert2016,grigoriev2016}. Overall, the localization becomes weaker as \R\
increases.
   %!TEX root=../main.tex
\section{Linear model of spatial decay}\label{sec:model}

\subsection{Mathematical formulation of the model}\label{subsec:method}

The exponential decay observed at the tails suggests that these can be
modelled with the linearised Navier-Stokes equations (LNSE). Following Gibson
\& Brand \cite{brand2014}, we look for normal mode solutions of the form
\begin{equation}\label{eq:ansatz}
   \begin{gathered}
      \vc{u}=\vc{\tilde{u}}(r)\exp\left[im\theta+\mu(z-ct)\right],\\
      p=\tilde{p}(r)\exp\left[im\theta+\mu(z-ct)\right],
   \end{gathered}
\end{equation}
where $m$ is the azimuthal wave number dominating at the tail, $\mu$ the
spatial decay rate at the tail and $c$ the group velocity at which the
localized solution (wave packet) travels in the axial direction. The latter
is not to be confused with the phase velocity of individual waves within the
solution (envelope). Note that $\mu$ is generally complex in a spatial
setting. Its real part describes the spatial attenuation (decay rate) and its
imaginary part the spatial modulation of the localized solution fronts.
Moreover, note that equation~\eqref{eq:ansatz} describes the tails in a
reference frame moving with the group velocity $c$. Although strictly
speaking this equation is only valid for relative equilibria (see
\cite{grigoriev2016}), the temporal variation is negligible at the tails of
our relative periodic orbits. Hence equation~\eqref{eq:ansatz} is used here
to model the spatial decay rates far away from their core.

Inserting ansatz \eqref{eq:ansatz} into the dimensionless LNSE gives:
\begin{equation}\label{eq:ansatz_nse}
   \begin{gathered}
      -\mu c\vc{\tilde{u}} = 
       (r^2-1)\mu\vc{\tilde{u}} 
       + 2r\tilde{u}_r\vc{\hat{z}}
      - \left[ \begin{array}{c}
         \partial\tilde{p}/\partial r\\
         im\tilde{p}/r \\
         \mu\tilde{p}
      \end{array} \right] 
       + \frac{1}{\R}L\vc{\tilde{u}}\\
       (\frac{1}{r} + \pp{}{r})\tilde{u}_r
       + \frac{im}{r}\tilde{u}_\theta + \mu\tilde{u}_z=0\textrm{,}
   \end{gathered}
\end{equation}
where
\[
L = \left[\begin{array}{ccc}
       \tilde{D} - \frac1{r^2} & -\frac{2im}{r^2} & 0\\
       \frac{2im}{r^2} & \tilde{D} - \frac1{r^2} & 0 \\
       0 & 0 & \tilde{D} \\
    \end{array}\right]
\]\\
with 
\[
\tilde{D} = -\frac{m^2}{r^2} +  \mu^2 + 
\frac{1}{r}\frac{\partial}{\partial r} + \frac{\partial^2}{\partial r^2}.
\]
Rearranging \eqref{eq:ansatz_nse} with respect to $\mu$ one obtains
the following quadratic eigenvalue problem (EVP):
\begin{equation} \label{eq:evp}
   (\mu^2 A_2 + \mu A_1 + A_0) 
   \left[\begin{array}{c}\tilde{u}_r\\\tilde{u}_\theta\\\tilde{u}_z\\\tilde{p}
   \end{array}\right] = 0,
\end{equation}
where
\[ A_2 =\dfrac{1}{\R}
\left[\begin{array}{cccc}
1 & & & \\ & 1 & & \\ & & 1 & \\ & & & 0
\end{array}\right],
\]

\[ A_1 =
\left[\begin{array}{cccc}
c+r^2-1 & & & \\ & c+r^2-1 & & \\ & & c+r^2-1 & -1\\ & & 1 & 0
\end{array}\right],
\]

and

\[ A_0 = \dfrac{1}{\R}
\left[\begin{array}{cccc}
 	\tilde{D} - \frac1{r^2} & -\frac{2im}{r^2} & 0 & -\R\pp{}{r} \\ 
 \frac{2im}{r^2} &  	\tilde{D} - \frac1{r^2} & 0 & -\R\cdot im/r \\
 2\R\cdot r & 0 & 	\tilde{D} & 0 \\ 
 \R\left(\frac{1}{r} + \pp{}{r}\right) & \R\cdot\frac{im}{r} & 0 & 0
\end{array}\right]
\]
with \[\tilde{D}=-\frac{m^2}{r^2} + \frac{1}{r}\pp{}{r} + \pp{^2}{r^2}.\]

The radial derivatives were discretized with a spectral method at
Chebyshev collocation points. In order to reduce clustering of grid
points near the origin (where the solution is smoother), the
differentiation matrices were computed over the interval [-1,1] using
$N=200$ points. The derivatives on (0,1] were obtained by ``quotienting''
out the symmetry of the 2-to-1 map from ($r-\theta$) to $(x,y)$ in this 
representation using the appropriate parities, respectively \cite
{tref2000}. 

The quadratic EVP \eqref{eq:evp} can be linearized analogous to the
reduction of a second-order ODE to first-order, namely by replacing it
with a \emph{linear} system with twice as many unknowns and equations \cite
{tisseur2001}. Here, we choose the so-called ``first companion form'' (see 
\cite{tisseur2001}) by making the substitution $\vc{y}=\mu \vc{x} \equiv \mu
[\vc{\tilde{u}}, \tilde{p}]^T$. This yields the generalized eigenvalue
problem
\begin{equation}
   \left[\begin{array}{cc}
      0 & I \\ -A_0 & -A_1
   \end{array}\right] 
   \left[\begin{array}{c}
      \vc{x} \\ \vc{y}
   \end{array}\right] = 
   \mu \left[\begin{array}{cc}
   I & 0 \\ 0 & A_3
      \end{array}\right]
   \left[\begin{array}{c}
      \vc{x} \\ \vc{y}
   \end{array}\right],
\end{equation}
which is subsequently solved with QZ-factorization (generalized Schur
decomposition, see \cite{moler1973}).

\subsection{Model results}
\label{subsec:full_mod}

\begin{figure}
   \includegraphics[width=\linewidth]{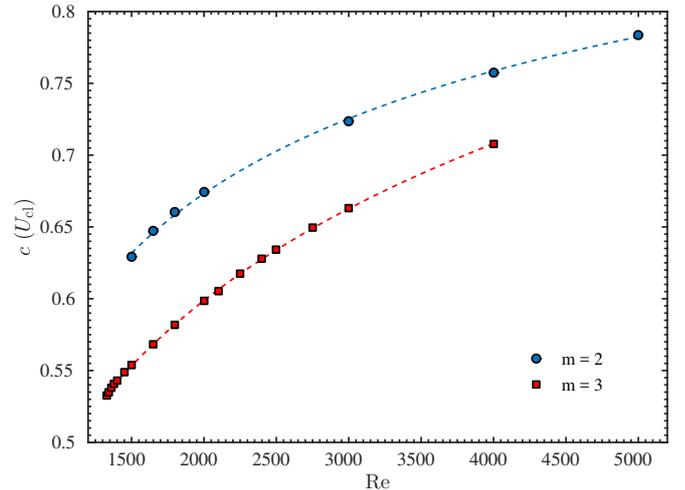}
   \caption{$\R$-dependence of the group velocity $c$ with which the
     localized solutions LB$_2$ and LB$_3$ are advected downstream.}
   \label{fig:c_Re}
\end{figure}

For given Reynolds number \R, azimuthal wave number $m$ and group velocity
$c$, positive (negative) eigenvalues of EVP \eqref{eq:evp} give
an approximation for the decay rate at the upstream (downstream) tail of a
localized solution. The associated eigenvectors approximate the velocity
profiles at the tails. 

The model predictions for the tails are computed as follows. First, the
Reynolds number is fixed and the group velocity $c$ is determined from the DNS.
Fig.~\ref{fig:c_Re} shows the evolution of $c$ as a
function of Reynolds number. Close to their saddle-node bifurcation points,
LB$_2$ and LB$_3$ travel slightly faster than the mean flow speed
$\overline{U}=\SI{0.5}{\Uc}$ and the differences grow slowly as \R\
increases. 

Second, the azimuthal wave number is determined from the velocity profiles.
As shown in fig.~\ref{fig:structure_fronts}, the upstream tail of LB$_3$
(LB$_2$) is dominated by a $m=6$ ($m=4$) rotational symmetry for all
components, whereas their downstream tails are predominantly axisymmetric for
$u_r$ and $u_z$, but feature $m=3$ ($m=2$) in $u_\theta$.  Thus, one needs to
consider both $m=0$ and $m=3$ ($m=2$) separately in the model for the
downstream tail. The case $m=0$ actually decouples the azimuthal velocity
from the other equations in EVP \eqref{eq:evp} leaving only the diagonal
block of the matrices $A_i$. Since their sum is not singular,
$\tilde{u}_{\theta}=0$ is obtained, which is not observed
(fig.~\ref{fig:structure_fronts}c) but consistent with the fact that $\tilde
{u}_{\theta}$ decays with a different rate than predicted by the axisymmetric
mode for $\tilde{u}_r$ and $\tilde{u}_z$. Moreover, the mean azimuthal
velocity (which is the $m=0$ mode) has to vanish because our localized
solutions are reflection symmetric, which precludes a mean rotation.

\begin{figure}
\includegraphics[width=\linewidth]{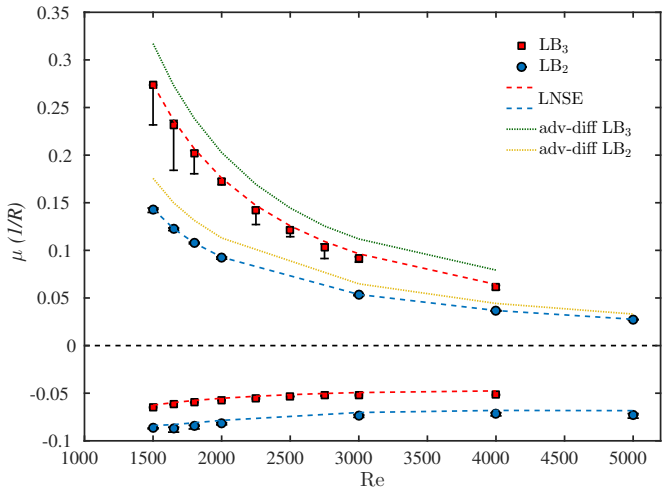}
\caption{
   Axial decay rate $\mu$ of $u_z$ for the localized solutions
   as a function of Reynolds number. Circles (LB$_2$) and squares (LB$_3$)
   denote DNS results. 
   Dashed lines denote the decay rates obtained from
   EVP~\eqref{eq:evp}. The downstream fronts have $m=0$, the upstream
   fronts $m=4$ for LB$_2$ and $m=6$ for LB$_3$. 
   Dotted lines show the result of the advection-diffusion equation \eqref
   {eq:adm} for the upstream front (see \ref{subsec:reduced_mod}).
}
\label{fig:fullmod}
\end{figure}

Figure~\ref{fig:fullmod} compares the decay rate $\mu$ of the
streamwise velocity disturbance obtained by exponential fits to $u_z$ at the
tails of the solutions (square/circular markers), to the prediction
based on the LNSE (dashed lines). The agreement is excellent, which
confirms the validity of the model.
\begin{figure}%\vspace{0.75mm}
\begin{tabularx}{\linewidth}{cc}
      (a) & (b) \\
      \begin{overpic}[width=0.5135\linewidth]{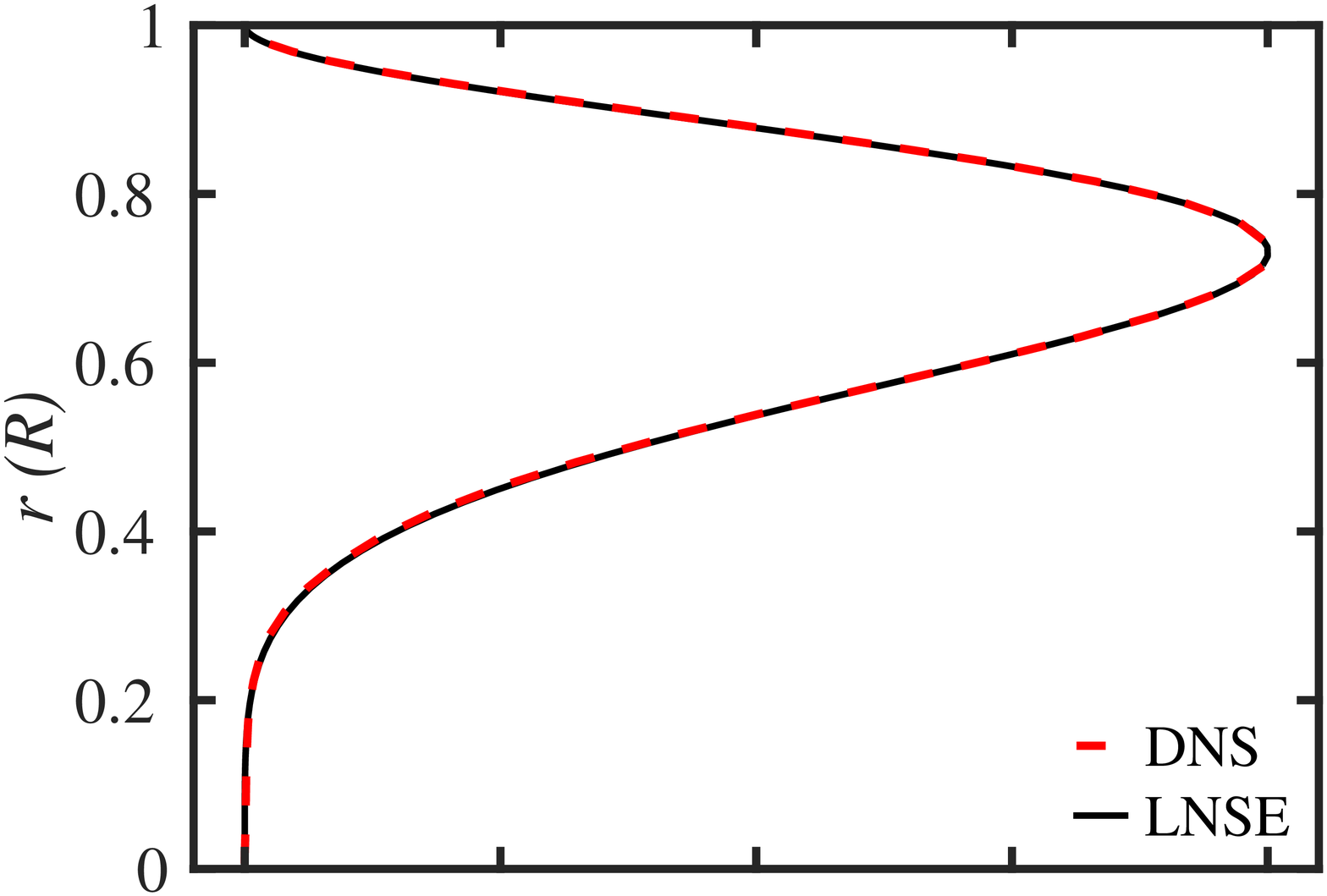}
      \put(132,1.4) {\includegraphics[width=0.446\linewidth]
      {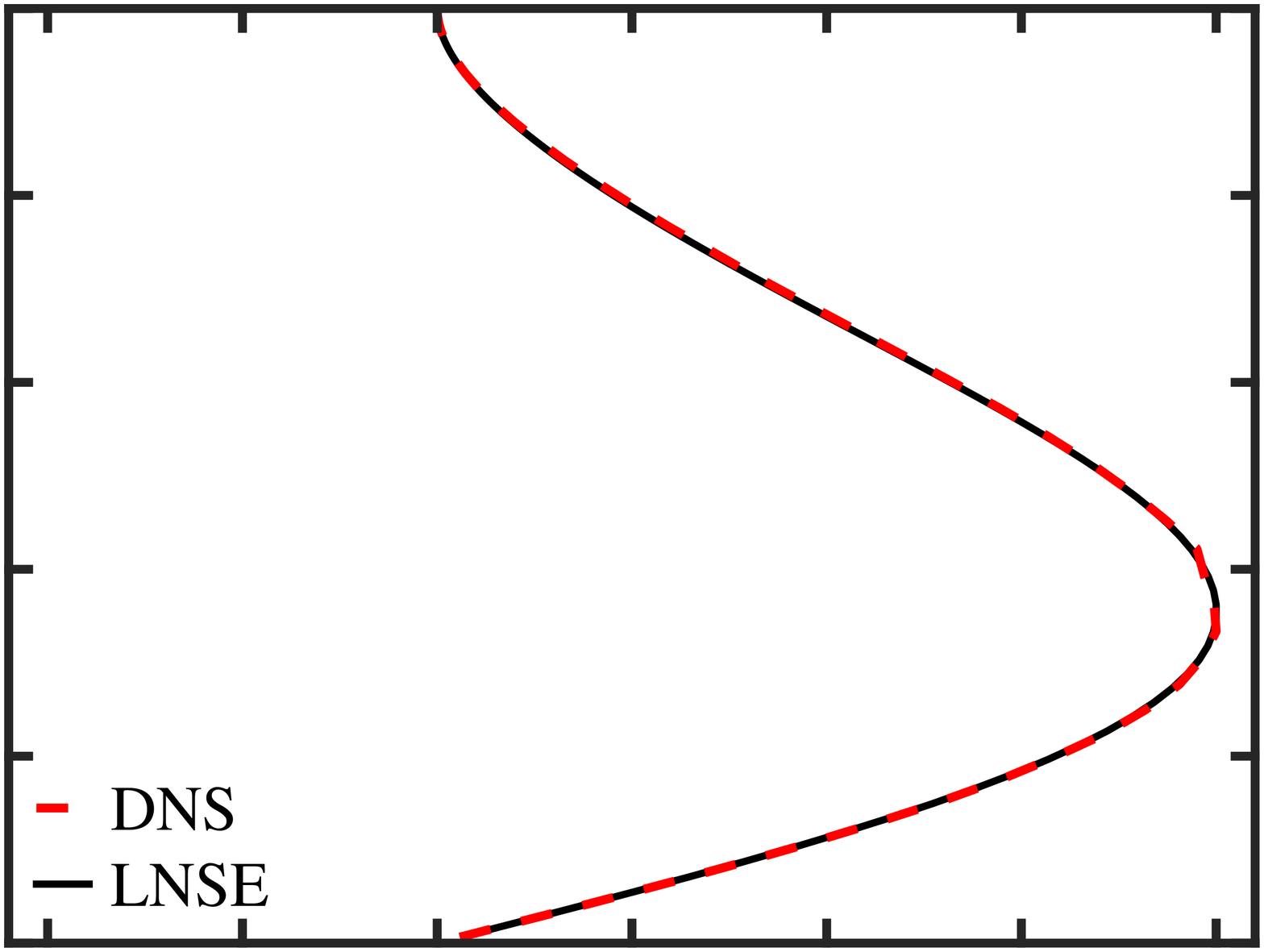}}
      \end{overpic} & \\
      (c) & (d) \\
      \begin{overpic}[width=0.5135\linewidth]{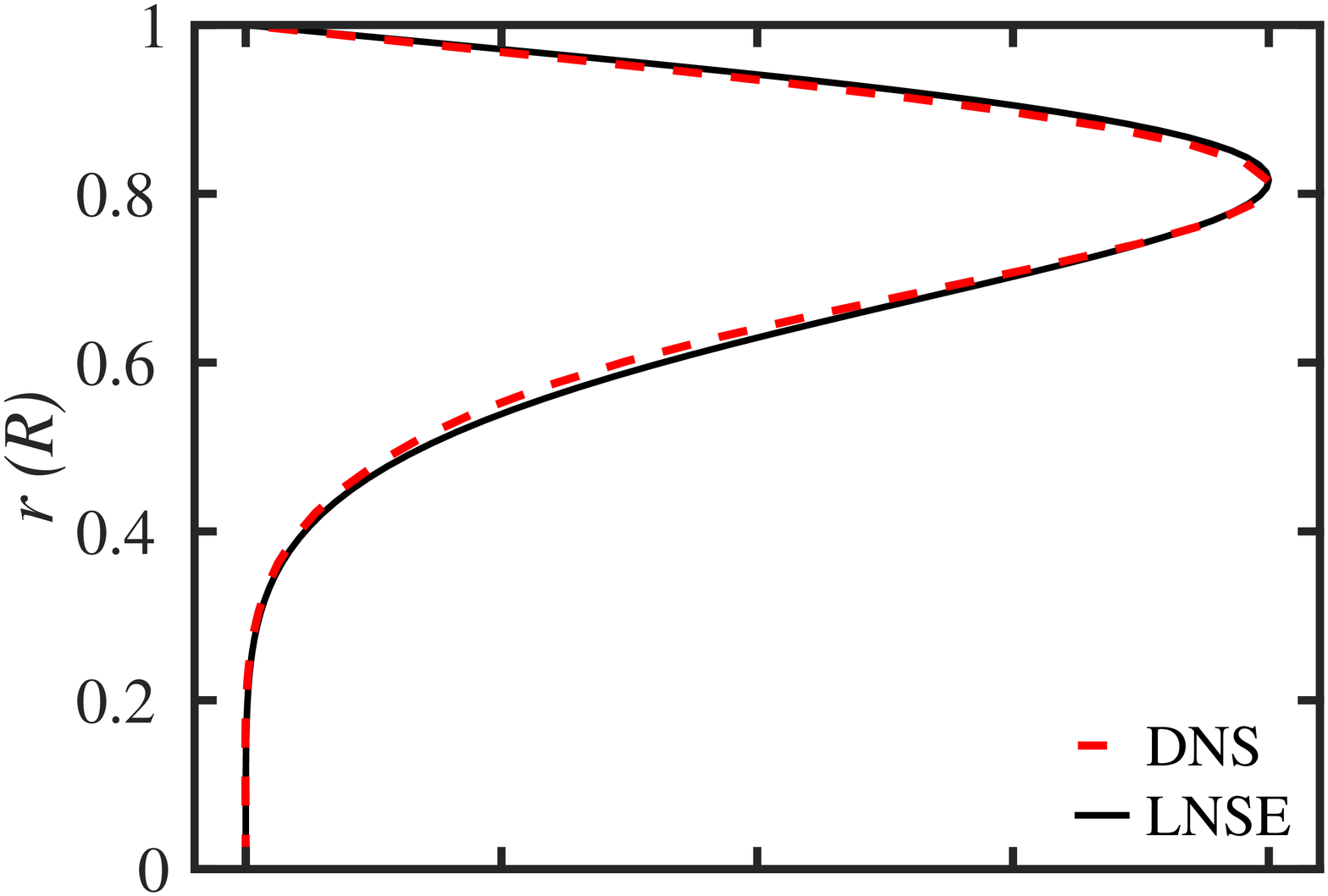}
      \put(132,1.4) {\includegraphics[width=0.445\linewidth]
      {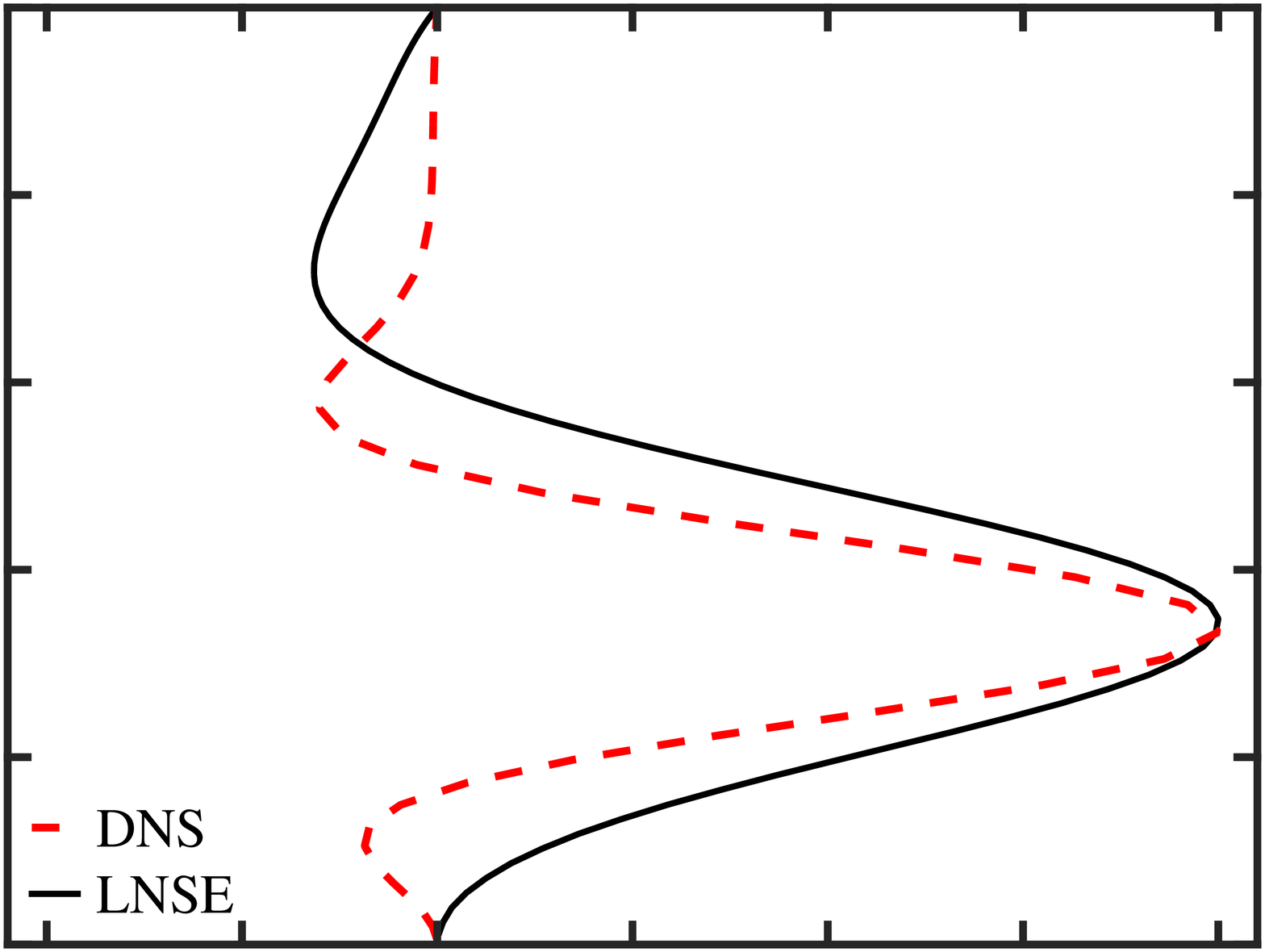}}
      \end{overpic} & \\
      (e) & (f) \\
      \begin{overpic}[width=0.5135\linewidth]{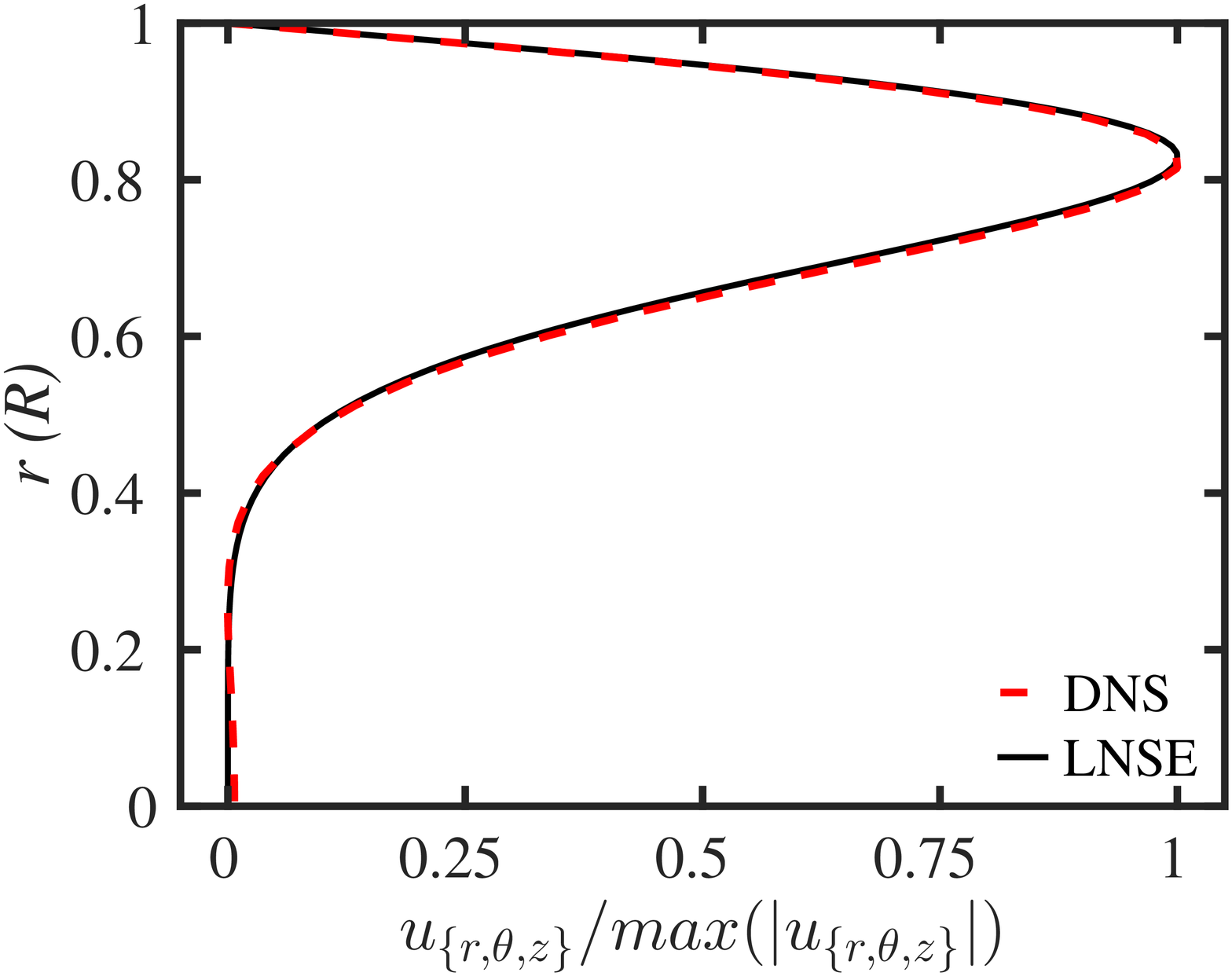}
      \end{overpic} &      
      \begin{overpic}[width=0.452\linewidth]{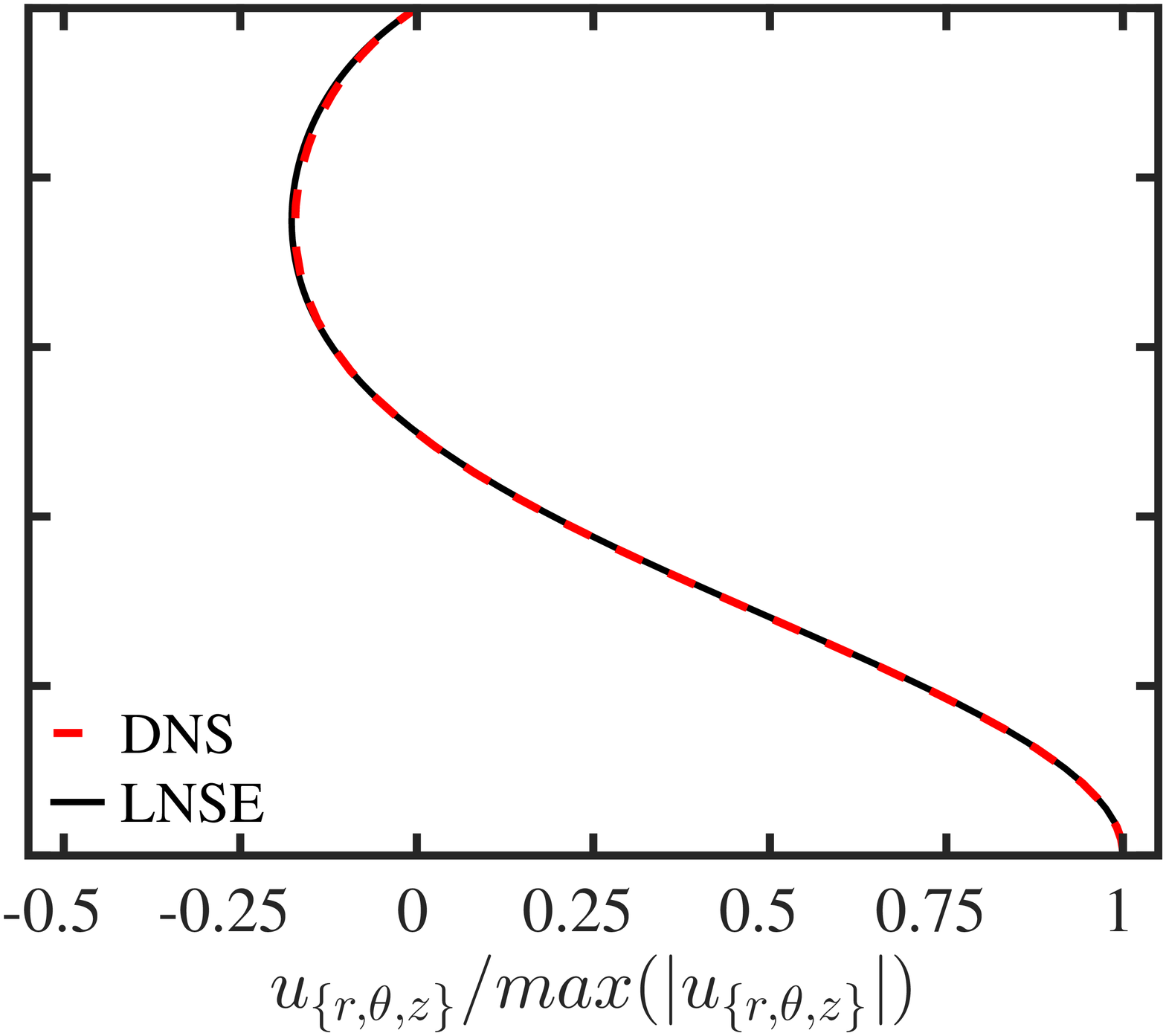}
      \end{overpic} 
\end{tabularx}
\caption{Comparison of radial profiles of (a, b) $u_r$, (c, d) $u_{\theta}$,
(e, f) $u_z$ obtained from DNS (dashed) and the LNSE (solid) for
LB$_3$ at $\RE{3000}$. The left hand side profiles are located up-, the
right hand ones downstream.}
\label{fig:profs_m3}
\end{figure}
The radial velocity profiles $\vc{\tilde{u}}(r)$ obtained from the model are
compared to the DNS data in fig.~\ref{fig:profs_m3} for LB$_3$. The agreement
of the upstream eigenvectors (model $m=6$) with DNS is very good. At the
downstream tail, the axisymmetric ($m=0$) model result for $\tilde{u}_r$ and
$\tilde{u}_z$ match the DNS result very well, too. To obtain a model
prediction for the azimuthal velocity, we solve the LNSE with $m=3$ as
suggested by fig.~\ref{fig:structure_fronts}c. This yields
$\Re[\mu]=\SI{-0.195}{\Lc^{-1}}$, whereas the decay rate obtained from DNS is
\SI{-0.25}{\Lc^{-1}} in both cases. 
However, the magnitude of $\tilde{u}_\theta$ is very small. Note also that
its decay is modulated in space (see figs.~\ref{fig:norm_lengths}b, \ref
{fig:structure_fronts}c) and this is correctly predicted by the model with
$\Im[\mu]\ne 0$. In all other cases, the imaginary part of $\mu$ is zero,
consistent with the absence of modulations in the spatial decay. The results
for LB$_2$ are very similar to those of LB$_3$ and hence not shown here.

\subsection{Contribution of terms to the spatial decay at tails}
\label{subsec:reduced_mod}

The decay rate of the streamwise tails of certain localized solutions in
Couette \cite{brand2014} and channel \cite{zammert2016} flows can be
accurately modeled with a single equation for the streamwise velocity
component of the disturbance. These authors compare the contributions of all
terms of the streamwise momentum conservation equation at the solution tails,
and find that three terms dominate: linear advection of the disturbance by
the basic laminar flow, and diffusion of momentum in the spanwise and 
wall-normal directions. The model resulting from consideration of only these
three terms is an advection-diffusion equation, which in pipe flow takes the
following form
\begin{equation}\label{eq:adm}
 -\left(c + r^2 -1\right) \mu\tilde{u}_z = \frac{1}{\R}\tilde{D} \tilde{u}_z.
\end{equation}

\begin{figure} 
  \includegraphics[width=\linewidth]{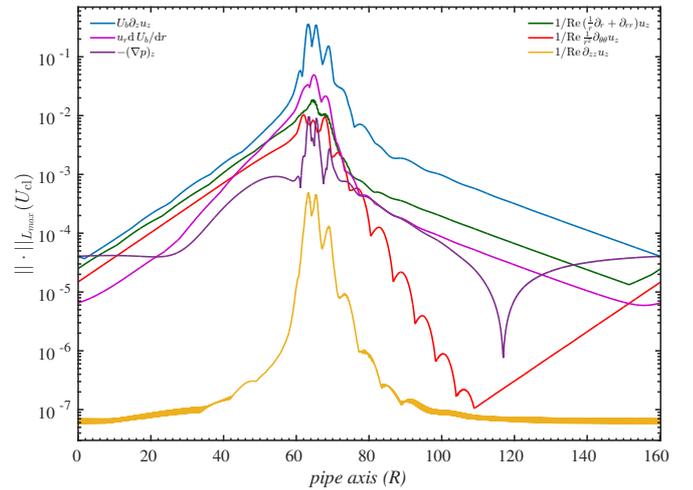}
  \caption{Axial profiles of the infinity norm (maximum) for each
    individual term in the axial component of the linearized
    Navier-Stokes equations at $\RE {3000}$ for LB$_3$.  Profiles are
    very similar for LB$_2$ and hence not shown here.}
\label{fig:terms}
\end{figure}

We computed the decay rates from \eqref{eq:adm} and found that they disagree
with those of the DNS and full LNSE, as also observed for spatially localized
modulated Tollmien-Schlichting waves in channel flow \cite{grigoriev2016}.
However, the upstream rates and eigenvectors $\tilde{u}_z$ computed from the
advection-diffusion model are at least comparable to the full simulation
(fig.~\ref{fig:fullmod}), whereas the downstream rates are utterly false
(hence not shown in fig.~\ref{fig:fullmod}). We assessed the reasons
underlying the failure of \eqref{eq:adm}, by analyzing individual
contributions of all terms in the streamwise momentum conservation equation
for LB$_3$ (see Fig.~\ref{fig:terms}, the relative contributions are similar
for LB$_2$ and hence not shown here). The diffusive term $\pp{^2}{z^2}$
(yellow) is much smaller than the in-plane contributions to $\nabla^2$ and
the other terms. In fact, setting $A_2=0$ and solving the resulting linear
eigenvalue problem does not have any effect on the decay rates, confirming
that axial diffusion can be neglected. The pressure gradient (violet) in the
upstream and downstream tails approaches its small and constant value in the
surrounding base flow. This value is a consequence of the periodic boundary
conditions and further decreases in longer pipes and with higher $\R$.

The relative contribution of the lift-up term $u_r\pp{U_\mathrm{b}}{r}$
(magenta) was found to be larger here than for the solutions of Zammert and
Eckhardt \cite{zammert2016} in channel and Brand and Gibson \cite{brand2014}
in Couette flows. In our solutions, the lift-up term is of similar magnitude
as the in-plane diffusion. This suggests that the absence of the lift-up term
in the simple advection-diffusion model \eqref{eq:adm} is responsible for its
failure. We gauged the role of the lift-up term by solving EVP~\eqref{eq:evp}
without it (\ie\ by omitting $2r$ in $A_0$). The corresponding upstream decay
rates are nearly identical to those from the advection-diffusion equation,
whereas the downstream rates  deviate strongly from the DNS (to a similar
degree as the advection-diffusion equation).

These findings indicate that the lift-up term plays a key role in the decay
of the tails in the pipe. For certain solutions of Couette and channel flows,
the main coupling of the streamwise velocity with the other components is via
the mass-conservation equation resulting in a very small wall-normal
velocity, which does not influence the axial decay rate (see the vanishing
value of $v$ in the tails in fig.~$2$a of \cite {zammert2016}). In pipe flow,
however, one cannot neglect $u_r$ in the tails and the radial momentum
equation is strongly coupled to the axial one via $u_r\pp{U_\mathrm{b}}{r}$.
Hence it is not possible to formulate an accurate single-equation model for
the decay of the streamwise velocity at the tails of LB$_2$ and LB$_3$,
exactly as for spatially localized modulated Tollmien-Schlichting waves in
channel flow \cite{grigoriev2016}.

\subsection{Application to a turbulent puff}

\begin{figure}
      \includegraphics[width=\linewidth]{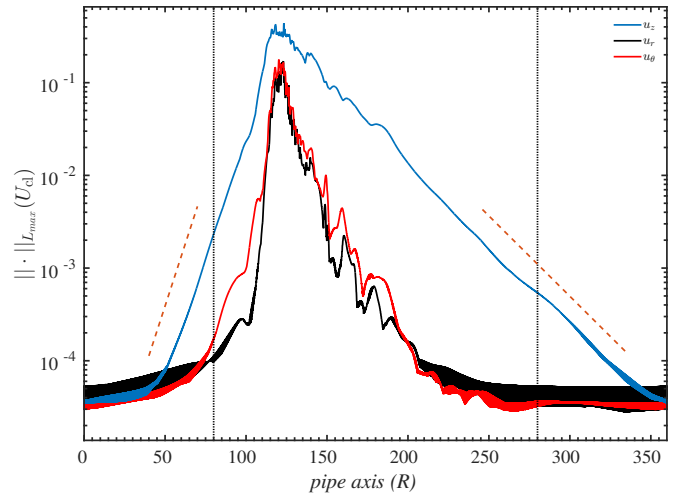}
      \caption{Axial profiles of infinity norm of the three
        velocity components for a turbulent puff at $\RE{2000}$ in a
        pipe of \SI{360}{\Lc} in length. The data are from Song
        \etal \cite{song2017}. The orange dashed lines show the spatial
        decay rates obtained from the model, where $m=2$ and $m=0$
        with $c/U_\mathrm{cl}=0.5$ were used to solve EVP
        \eqref{eq:evp} for the upstream and downstream tails,
        respectively. The dotted vertical lines indicate the locations of the
        velocity profiles shown in fig.~\ref{fig:profs_puff}.}
\label{fig:norm_lengths_puff}
\end{figure}
\begin{figure}
  \includegraphics[width=0.52\linewidth]{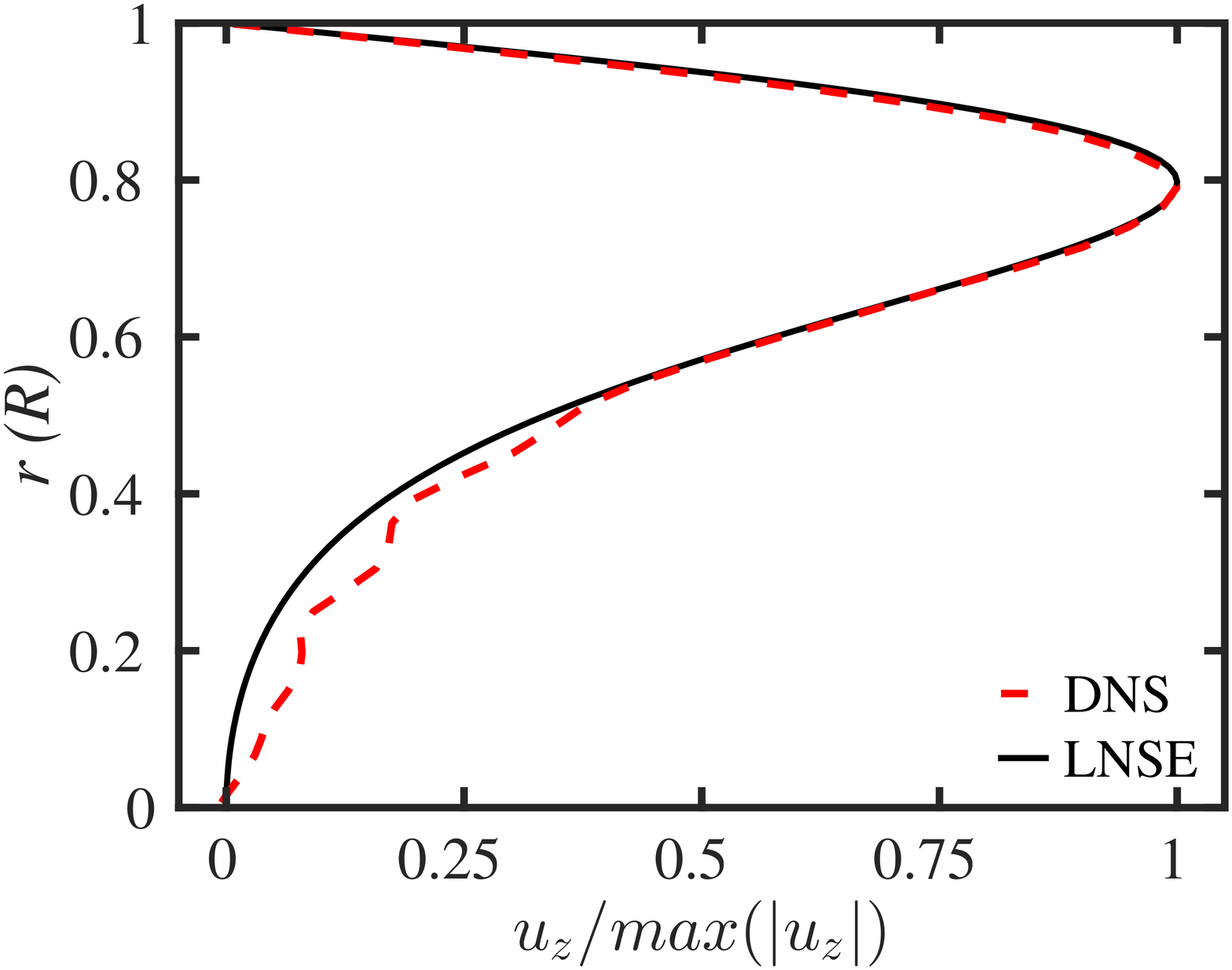}
  \includegraphics[width=0.457\linewidth]{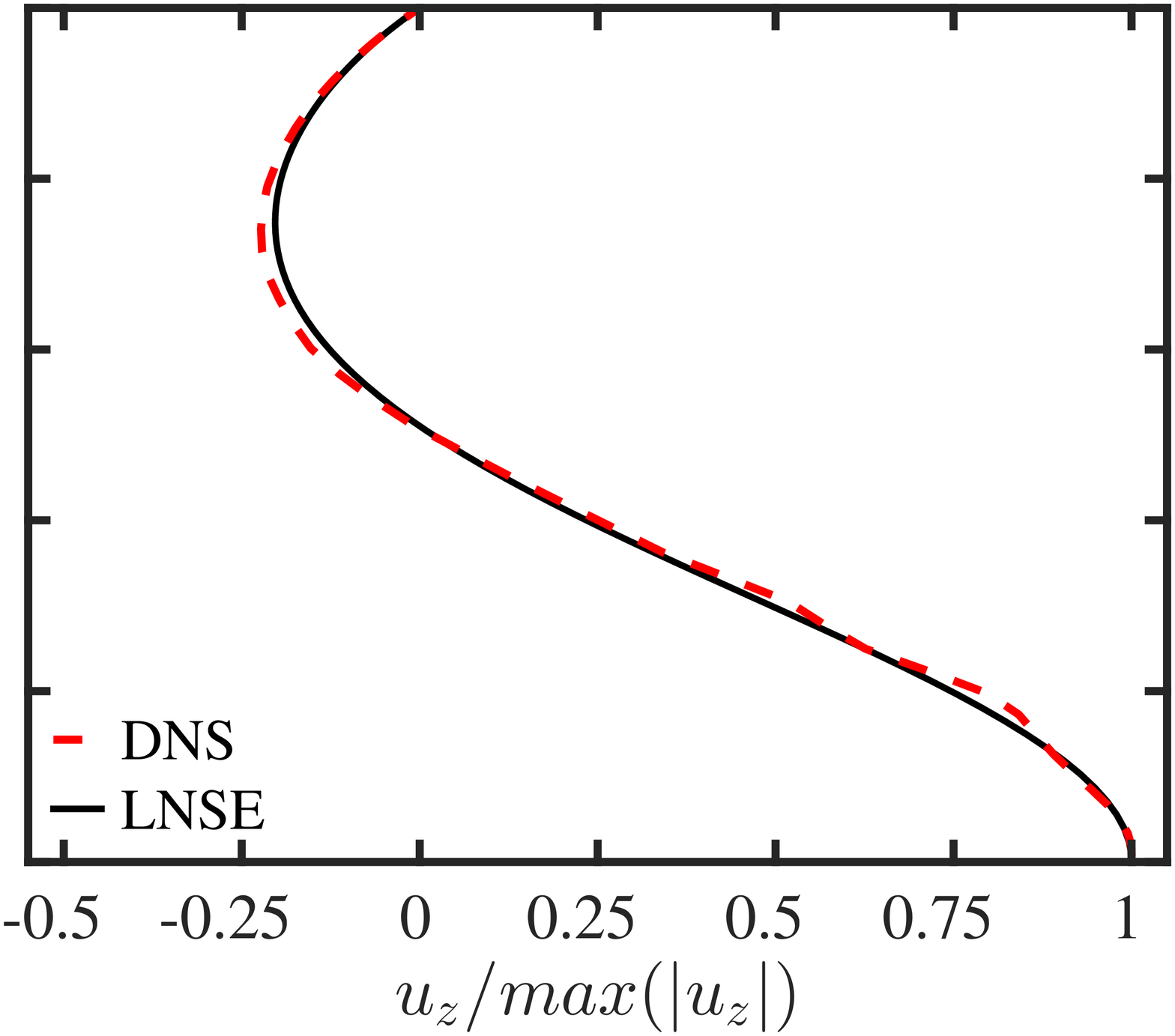}
  \caption{Comparison of radial profiles of $u_z$ obtained from DNS (dashed) and the full
    model (solid) for a puff at $\RE{2000}$. The left hand side profile
    is located upstream, the right hand one downstream. In the model,
    $m=2$ and $m=0$ with $c/U_\mathrm{cl}=0.5$ were used to solve
    EVP \eqref{eq:evp} for the upstream and downstream tails,
    respectively.}
\label{fig:profs_puff}
\end{figure}
% We should ask Baofang, if when the time-step is reduced or the resolution
% is increased, we can get cleaner results. At the moment we go to numerical
% noise at 10^-4 already, which is not so great. You can try yourself as
% well. First halve dt, if this does not help then in addition you increase
% the resolution. This is not urgent, but worth doing at least for your PhD
% or for when we get the reviews of the paper

Models based on LNSE have been so far applied to describe the tails of exact
coherent solutions.  Mellibovsky \etal \cite{fer2009} showed that the tails
of an edge state and a turbulent puff at $\R=1900$ decay exponentially,
suggesting that LNSE may correctly describe their spatial decay. We here
applied the LNSE to a turbulent puff at $\RE{2000}$ (see
fig.~\ref{fig:norm_lengths_puff}), which propagates at exactly the mean speed
$c/U_\mathrm{cl}=0.5\pm0.00025$ \cite{avila2011}.  Using this value of $c$
yields
$\mu=\SI{0.13}{\Lc^{-1}}$ and $\mu=\SI{-0.0395}{\Lc^{-1}}$, for the upstream
($m=2$) and downstream ($m=0$) front, respectively, which is in excellent
agreement with the results from exponential fits to $u_z$ at the tails of the puff.
Similarly the velocity profiles from the model match those of the DNS at the
tails as shown in fig.~\ref{fig:profs_puff}.
% I suppose that you are using c=1, but don't really know
% This is based on one snapshot, what happens when we use more snapshots? Same m=2 adjusts best upstream?
% Extra figure needed (may be for the paper, for sure for the PhD thesis):
% Show max |u_z(m,z)| for m=0,1,2,3,4 (four curves), to verify that m=0 and m=2 dominate at the downstream and upstream tails
% Try a puff at a different Re (ask Baofang what he has, I would suggest Re=1900,2100,2200
% c changes with Re and we can also see if the same m dominates upstream or not.

   %!TEX root=../main.tex
\section{Discussion}

Localized exact coherent structures in pipe flow exhibit exponential
localization far away from their active core. This allows for accurate models
of the decay based on the LNSE, as in the cases of Couette
\cite{gibson2014,brand2014} and channel flow \cite{zammert2016,
grigoriev2016}. The solution of the resulting spatial eigenvalue problem
yields two decay rates of different sign for the velocity disturbance
at the upstream ($\mu>0$) and downstream tails ($\mu<0$), as a function of
Reynolds number \R, azimuthal wave number $m$ and group velocity $c$.  

The localized solutions investigated here are relative periodic orbits and
have either two- or three-fold rotational symmetry and are reflectional
symmetric. Their upstream tails feature four/six streaks and all three
components of the velocity disturbance decay at the rates as in the
model for $m=4/6$. Their downstream tails are predominantly axisymmetric and
consist of large scale meridional circulation $(u_r,u_z)$, whereas $u_\theta$
presents a two-/three-fold symmetry, is much smaller and decays at a much
faster rate. The model accurately predicts the decay rates for $(u_r,u_z)$
using $m=0$ at the downstream tails, but the decay of $u_\theta$ is only qualitatively  
recovered even when $m=2,3$ are used in the model.

The decay rate of some solutions in Couette \cite{brand2014} and channel
\cite{zammert2016} flows can be accurately modeled with a single advection-
diffusion equation for the streamwise velocity disturbance. Interestingly, in
this equation the Reynolds number and decay rate appear only through the
combination $\mu\R$, so that one may expect $\mu$ to decrease as $1/\R$.
However, the propagation speed of the structures was found to increase with
\R\ in channel flow leading to a faster decrease of $\mu$ with \R\ upstream
and a slower, nearly constant increase downstream. For spatially localized
modulated Tollmien-Schlichting waves in channel flow \cite{grigoriev2016},
and for our solutions, a simple advection-diffusion equation cannot reproduce
the results from the full LNSE. We here showed that the lift-up term
$u_r\pp{U_b}{r}$ is significant enough so that it cannot be neglected. This
term couples the radial and axial momentum equations and so it is no longer
possible to retain a single equation model for the decay rate of the
streamwise velocity disturbance. Despite the key role of the lift-up term, we
still found that the scaling of the decay rates with \R\ corresponds to what
one would expect from the advection-diffusion equation \eqref{eq:adm} once
the dependence on $c(\R)$ is taken into account, similar to channel flow
\cite{zammert2016,grigoriev2016}.

The LNSE were also applied to data from DNS of localized turbulence. Here a
turbulent puff at $\R=2000$ was analyzed. The spatial decay rates obtained
from the model with $c/U_\mathrm{cl}=0.5$, and $m=0,2$ for the downstream and
upstream tails, were found to be in excellent agreement with the DNS data.
This confirms the validity of the model for real turbulent patches and
emphasizes the crucial interdependence between propagation speed of a
localized structure and the spatial decay rate at its tails. Note that 
the dependence of $\mu$ on the group velocity $c$ suggests that in the transition
from localized puffs to expanding slugs the localization rate must change accordingly.

   \begin{acknowledgments}
     Support from the Deutsche Forschungsgemeinschaft (DFG) through grant
     FOR 1182 and computing time from the ``Regionales Rechenzentrum
     Erlangen (RRZE)'' are acknowledged.  This research was supported in
     part by the National Science Foundation under Grant No.~NSF
     PHY-1125915. S.~Z. acknowledges financial support by Stichting FOM/NWO-I.
   \end{acknowledgments}

   % \bibliography{paper_tails}

\end{document}